\documentclass[acmtog]{acmart}

\AtBeginDocument{%
  \providecommand\BibTeX{{%
    \normalfont B\kern-0.5em{\scshape i\kern-0.25em b}\kern-0.8em\TeX}}}

\setcopyright{acmlicensed}
\copyrightyear{2023}
\acmYear{2023}
\acmDOI{XXXXXXX.XXXXXXX}

\acmVolume{1}
\acmNumber{1}
\acmArticle{1}
\acmMonth{11}

\begin{document}

\title{Enhancing EmoBot: An In-Depth Analysis of User Satisfaction and Faults in an Emotion-Aware Chatbot}

\author{Taseen Mubassira}
\email{taseen123buet@gmail.com}
\affiliation{
  \institution{Bangladesh University of Engineering and Technology}
  \city{Dhaka}
  \country{Bangladesh}
}

\author{Mehedi Hasan}
\email{mhasan912@gmail.com}
\affiliation{
  \institution{Bangladesh University of Engineering and Technology}
  \city{Dhaka}
  \country{Bangladesh}
}

\author{A. B. M. ALIM AL ISLAM}
\email{razi_bd@yahoo.com}
\affiliation{
  \institution{Bangladesh University of Engineering and Technology}
  \city{Dhaka}
  \country{Bangladesh}
}

\renewcommand{\shortauthors}{Taseen, et al.}

\begin{abstract}
\large{ABSTRACT} \\
  The research community has traditionally shown a keen interest in emotion modeling, with a notable emphasis on the detection aspect. In contrast, the exploration of emotion generation has received less attention.This study delves into an existing state-of-the-art emotional chatbot, EmoBot, designed for generating emotions in general-purpose conversations. This research involves a comprehensive examination, including a survey to evaluate EmoBot's proficiency in key dimensions like usability, accuracy, and overall user satisfaction, with a specific focus on fault tolerance. By closely examining the chatbot's operations, we identified some noteworthy shortcomings in the existing model. We propose some solutions designed to address and overcome the identified issues.
\end{abstract}




\received{15 November 2023}
\received[revised]{}
\received[accepted]{}

\maketitle

\section{Introduction}
In recent years, there has been a growing emphasis on the study of social chatbots. Unlike older rule-based counterparts, modern chatbots, driven by deep learning, have shown significant improvements \cite{sutskever2014sequence}. In order to design a chatbot that provides a meaningful experience, we must first understand what expectations people have for this technology, and what opportunities are there for chatbots based on user needs \cite{zamora2017m}. As chatbots become more prevalent in areas like entertainment and customer service \cite{io2017chatbots}, the focus has shifted to making them more emotionally responsive and human-like. The goal is to enable chatbots to engage in empathetic conversations, often assuming the role of a social companion, thereby positively impacting the well-being of individuals \cite{skjuve2021my}. Therefore, adaptability to various scenarios and meeting user needs are crucial for a social chatbot's success.This study delves into Emobot, an emotional chatbot using the cognitive appraisal theory to generate emotions based on user responses \cite{ehtesham2024emobot}.  Our analysis involves an exploration of user experiences with this chatbot. We examine user experiences with Emobot, identifying areas for improvement in its emotion generation. Our goal is to detect and address these issues, proposing practical solutions to enhance the chatbot's performance. Our contribution in this light are as follows : 
\begin{itemize}
    \item Conducted a survey to assess user experience and subsequently performed a qualitative analysis on the gathered data.
    \item By examining the codebase and analysing user experince we have found some faults in the existing Emobot.
    \item Proposed different solution approaches according to the shortcomings.
\end{itemize}

\section{System Design}
\begin{figure}
  \includegraphics[width=\linewidth]{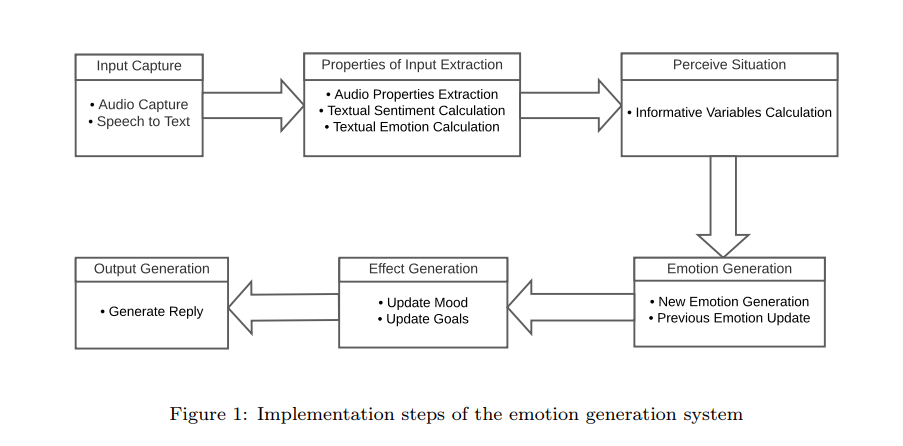}
  \caption{Implementation steps of the emotion generation system}
  \label{fig:boat1}
\end{figure} 

The current chatbot, Emobot, is structured according to a system design that consists of six steps, as illustrated in Figure 1.
\subsection{Overview of the steps : }
\textbf{Input capture:} Input can be captured in both audio or textual format. When audio serves as the primary input, it is converted into its corresponding text using a speech-to-text system. \\
\textbf{ Properties of input extraction:} From audio and text input, corresponding auditory and textual properties are extracted. \\
\textbf{Informative variables to evaluate an event (appraisal variables):} Evaluate multiple mappings between the 10 informative variables(based on cognnitive appraisal theory) and the captured input to find the most prominent one. \\
\textbf{Emotion Generation :} Emotions are generated based on informative variables that assess system events, resulting in the primary emotions: Joy, Sadness, Fear, Anger, and Surprise. \\
\textbf{Effect Generation:} The influence of a generated emotion extends to changes in mood states, personality traits, and the system's prior emotional condition. Mood states are characterized using the PAD model \cite{zhou2018multi}, while the system integrates personality traits according to the five-factor mode \cite{baranczuk2019five}. \\
\textbf{Output Generation:} The system produces a response, presented to the user in both audio and text formats.

\section{Methodology}
Following the analysis of the current Emobot, we conducted a survey to assess user satisfaction and system relevance. The survey involved 13 participants who provided input to the chatbot in both audio and textual formats. The average chat duration for the audio chat was 2 min and for the textual input it was 6 min. Upon gathering user responses for Emobot, they were asked to rate the current system by four metrics on a scale of 0-10 to evaluate the performance of Emobot in the eyes of users. Based on the information fed from different users, we calculated 4 metrics to base on the performance of EmoBot. These metrics were used for finding the shortcomings and faults in the current system which is discussed in the following section.

\begin{figure}[H]
  \includegraphics[width=\linewidth]{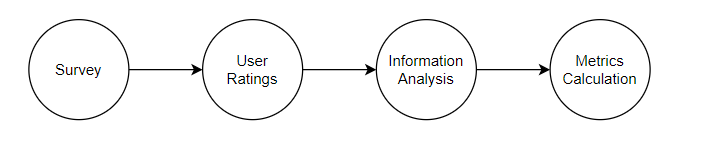}
  \caption{Flowchart of the work procedure}
  \label{fig:boat1}
\end{figure}

\subsection{Metrics}
The metrics we used for determining the performance of EmoBot are:

\begin{itemize}
    \item \textbf{Relevance}: It measures whether EmoBot was able to provide a response that aligned with the contest of user's prompt.
    \item \textbf{Satisfaction}: It evaluates whether the user was pleased with the response provided by EmoBot.
    \item \textbf{Accuracy}: It assesses whether EmoBot's reply conveyed the same emotion as the user's prompt.
    \item \textbf{Appropriateness}: It evaluates whether EmoBot's reply seemed fitting and suitable to the user.
\end{itemize}

\section{Findings and Analysis}
Our analysis with the survey results and the metrics calculated from the user experience, it can be seen that the average user satisfaction is around 3.8 in figure 3. 

\begin{figure}
  \includegraphics[width=\linewidth]{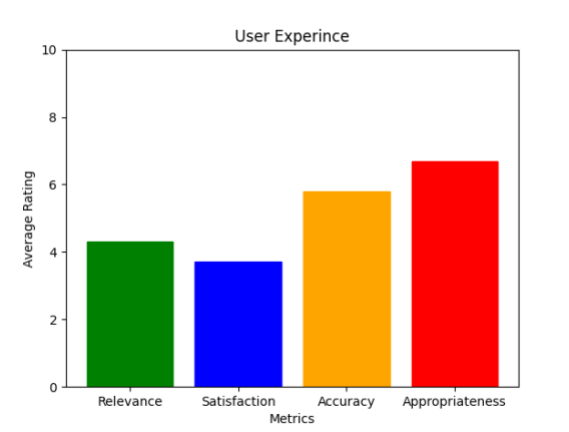}
  \caption{Average ratings on the metrics from the survey}
  \label{fig:boat1}
\end{figure}

Upon exploring the code of EmoBot and correlating it with our survey findings, our investigation unveiled a significant aspect, leading to the identification of certain faults in the current modeling approach. These are as follows: 
\begin{itemize}
    \item  All mapping inputs were derived from human inputs
    \item An interesting observation surfaced when scrutinizing the selected approach, which aligned with the methodology employed by the study's authors. However, this approach lacked a comprehensive explanation or research-backed justification for the specific numerical choices made
    \item Additionally, a critical issue came to light during the evaluation of EmoBot's performance—the bot  misidentified "surprise" as "joy." Further investigation uncovered a bug in the code responsible for this misclassification. It is shown in figure 4.
    \item Moreover, the bot exhibited a tendency to misjudge sentences with misleading primary emotions and complex emotions, as exemplified by the response to a statement like "I was excited to join the party but could not due to rain," where the bot incorrectly labeled the emotion as "joy" when a more suitable classification would be "sad."
\end{itemize}
\begin{figure}
  \includegraphics[width=\linewidth]{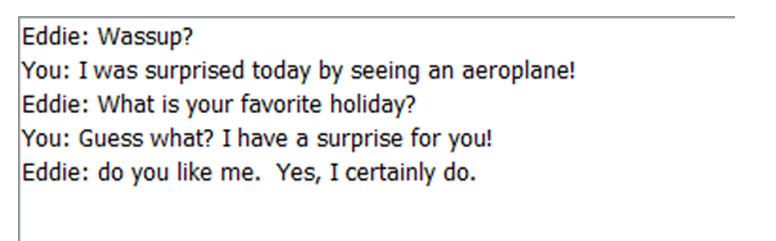}
  \caption{Emobot fails to recognize "surprise"}
  \label{fig:boat1}
\end{figure}

\begin{figure}
  \includegraphics[width=\linewidth]{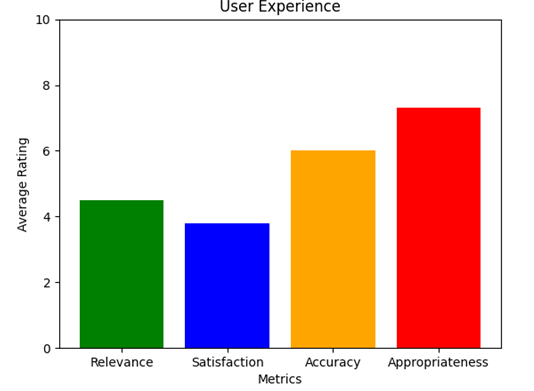}
  \caption{Average ratings on enhanced Emobot from the survey}
  \label{fig:boat1}
\end{figure}

\section{Finetuning Emobot}
After identifying and resolving a bug that frequently led Emobot to misinterpret "surprise" as "joy," adjustments were made to the human input mappings to align with the corrected functionality. Following this, extensive experimentation with various mapping values ensued. Following these efforts, a survey involving previous users was undertaken to evaluate Emobot's improved responsiveness to joy, surprise, and fear. The results indicated substantial enhancements in both relevance, which increased from 4.1 to 4.5, and appropriateness, rising from a score of 7 to 7.3. These improvements signify a significant advancement in Emobot's capacity to interact effectively with users, resulting in heightened relevance and appropriateness in its responses.

\section{Future Work and Proposed Solution}
In response to the identified faults, we have proposed the following solution approaches in the current EmoBot system :
\begin{itemize}
    \item Refine the bot's modelling
    \item Develop the ability to retain context from previous conversations for better relevance
    \item Implement a machine learning based approach to determine the mapping values
    \item Investigate advanced psychological theories instead of relying solely on cognitive appraisal theory 
\end{itemize}

\bibliographystyle{ACM-Reference-Format}
\bibliography{EnhancingEmobot}


\end{document}